**Between Light and Eye: Goethe's Science of Colour and the Polar Phenomenology of Nature**

Alex Kentsis[*]

**Abstract**

In his interviews with Eckermann in the 1820s, Goethe referred to his *Theory of Colours* as his greatest and ultimate achievement.[1] Its reception following publication in 1810 and subsequent reviews throughout the history of physical science did not reflect this self-assessment. Even Hermann von Helmholtz, who in part modeled his scientific work after Goethe's, initially thought that Goethe's poetic genius prevented him from understanding physical science.[2] Why did Goethe champion his *Farbenlehre* so ardently even years after it was dismissed by almost everyone else? In answering this question, this essay will attempt to add to the existing scholarship by considering Goethe's *Theory of Colours* in the context of his natural philosophy, and generalizing the variety of objectifications of the concepts invoked by his colour theory and their relationships to Goethe's epistemology and Newton's mechanics. In this fashion, I attempt to show that the reason for Goethe's self-assessment of his *Theory of Colours* is less enigmatic than appears from its examination solely as a work of physics. Rather, *Zur Farbenlehre* was the clearest expression of Goethe's most universal archetype—polarity of opposites— which bridged Goethe's conflicts with Kant's and Spinoza's epistemologies, and in an over-reaching way served as a cosmology underlying Goethe's art and his science.

---

[*] Mount Sinai School of Medicine, New York University, New York, New York, 10029, USA.

[1] J. P. Eckermann, *Conversations of Goethe*, (tr. J. Oxenford), London, 1930, 301-2.

[2] H. von Helmholtz, 'On Goethe's Scientific Researches', in *Science and Culture* (ed. D. Cahan), Chicago, 1995, 7.

**Goethe's Phenomenalism**

At least in its departure, Goethe's *Theory of Colours* was a reaction to Newton's *Opticks*. While Goethe recognized the popularity of Newtonian natural philosophy, he strongly objected to its widespread use. With respect to colour theory, Goethe believed that 'Newton had based his hypothesis on a phenomenon exhibited in a complicated and secondary state'.[3] He objected not so much to the abstract argumentation of rational mechanics, commenting that the 'mathematician would willingly cooperate with us, especially in the physical department of the theory',[4] but rather to the ground premise of Newton's approach. Goethe protested the destruction of phenomenal essence that occurred during the phenomenon's 'dissection into its component parts'.[5] As a consequence of this irreducibility,

> the scientific minds of every epoch have also exhibited an urge to understand living formations as such, to grasp their outward, visible, tangible parts in context, to see these parts as an indication of what lies within and thereby gain some understanding of the whole through an exercise of intuitive perception.[5]

In this way, Goethe sought to develop an understanding of objective, empirically accessible features of natural phenomena in their context in order to gain a subjective understanding of 'what lies within'.[5]

---

[3] J. W. von Goethe, *Theory of Colours*, (tr. C. L. Eastlake), Cambridge, 1994, iii.

[4] Goethe, op. cit. (3), lx.

[5] J. W. von Goethe, *Scientific Studies*, (tr. D. Miller), New York, 1988, 63.

**Newton's Reductionism**

In contrast, Newton's purpose of *Opticks* was 'not to explain the Properties of Light by Hypotheses, but to propose and prove them by Reason and Experiment'.[6] Thus, Newton began with eight definitions of common optical terms (ray, incidence, etc), followed by eight axioms which state the principles of geometric optics (reflection, refraction, etc). Proposition I stated that 'Lights which differ in Colour, differ also in Degrees of Refrangibility,' as proved using a card, half of which was painted red and the other half blue, placed against a black background, and viewed through a prism.[7] Having observed that the blue half of the card appeared lower than the red, Newton concluded that the colour blue 'suffer[ed] a greater Refraction' than the colour red.[8] Proposition II stated that white light 'consists of Rays differently Refrangible,' as proved by using a circular aperture and a prism to project a beam of light onto a wall, and viewing the cast colour spectrum.[9] The reduction of the phenomenon of colour to a mechanism inherent in light was summarized in the following definition from the *Opticks*:

> Colours in the Object are nothing but a Disposition to reflect this or that sort of Rays more copiously than the rest; in the Rays they are nothing but their Dispositions to propagate this or that Motion into the Sensorium, and in the Sensorium they are Sensations of those Motions under the Forms of Colours.[10]

---

[6] I. Newton, *Opticks*, New York, 1952, 1.

[7] Newton, op. cit. (6), 20.

[8] Newton, op. cit. (6), 21.

[9] 'The image … was coloured, being red at its least refracted end, and violet at its most refracted end, and yellow green and blue in the intermediate Spaces.' Newton, op. cit. (6), 26.

[10] Newton, op. cit. (6), 125.

Newton provided a mechanism for the creation of colour with its essential components being contained in light as an object, instead of functioning in the Sensorium of the subject, or instead of the interaction between light and its perceived and intuited effect. It is precisely this abstraction of concepts from intuition by Newton that Helmholtz attributed as the cause of Goethe's passionate rejection of the Newtonian doctrine of colour.[11] For Goethe, Newton's explanation was unacceptable because it failed to explain why this mechanism (of motion) was related to our particular modes of perception, and should give rise to colour at all.

**Goethe's Reverence of the Natural Object**

In contrast to Newton's, Goethe's natural philosophy was not interested in the decomposition of phenomena into their causal processes, either mechanical or genealogical. Rather, Goethe sought 'conditions under which phenomena appear; their consistent succession, their eternal return under thousands of circumstances [and] their uniformity and mutability'.[12] This 'mistrust of abstraction'[13] was an expression of Goethe's reverence of the natural object. Providing direction, this guide warned against the idealization of Nature by the natural philosopher who 'should be careful not to transform perceptions into concepts, concepts into words, and then treat these words as if they were objects'.[14]

---

[11] Helmholtz, op. cit. (2), 403.

[12] Goethe, op. cit. (5), 25.

[13] J. Barnouw, 'Goethe and Helmholtz: Science and Sensation', in *Goethe and the Sciences: A Reapraisal* (eds. F. Amrine, F. J. Zucker, H. Wheeler), Boston, 1987, 56.

[14] Goethe, op. cit. (3), liii.

The primacy of the natural object had a dual foundation. Both were what Goethe would have termed the historical determinants of his life. On one hand, the primacy of the natural object stemmed epistemologically from Goethe's Spinozistic monism in identifying Nature and God.[15] On the other hand, it may have been related politically to Goethe's aversion to the French Revolution, seeing it as a tragic and terrifying idealization of the social world by the body politic, and fearing an analogous idealization of the natural world by science.[16] The potentially deceptive nature of judgment saw its expression in Goethe's belief that the modern age tended to be scientifically inaccessible and abstruse. As he pointed out:

> We remove ourselves from common sense without opening up to a higher one; we become transcendent, fantastic, fearful of intuitive perception in the real world, and when we wish to enter the practical realm, or need to, we suddenly turn atomistic and mechanical.[17]

As such, sanctioning the priority of conceptual knowledge over the intuitive would have terrorizing consequences for understanding of the natural world, much like the unchecked pursuit of social ideals did during the French Revolution. Thus, Goethe wrote that he himself had initially accepted Newton's theory of white light until a particular experience brought him to his senses and to an understanding firmly grounded in intuition.[18]

---

[15] G. Floistad, 'Spinoza's Theory of Knowledge and the Part-Whole Structure of Nature', in *Spinoza on Knowledge and the Human Mind*, (ed. Y. Yovel), 2 vols., Leiden, 1994, i, 45-67.

[16] G. Brande, *Wolfgang Goethe*, (tr. A. W. Porterfield), 2 vols., New York, ii, 29-41.

[17] Goethe, op. cit. (5), 309.

[18] Barnouw, op. cit. (13), 67.

**Light and Eye: The Proximal Prism Experiment**

In addition to Goethe's natural philosophy being defined by its reverence of the natural object, it is just as critically characterized by its development of a holistic understanding of phenomena. In this way, it was related both to the psycho-physical parallelism of Spinoza, as well as to the dependence of intuition on the functions of the mind, as developed by Kant.

Thus, Goethe sought to capture the universal essence of the objective world in the particular way of a subjective world-view. Writing in the *Italian Journey*, Goethe remarked about wishing to make a journey to India, 'not for the purpose of discovering something new, but in order to view in [his] way what has [already] been discovered'.[19] Thus, for Goethe, as well as his colour theory, it was the way of seeing, and seeing in a very particular way, that organized the features of the objective experience of Nature.[20]

As a result, when Goethe saw that the prismatic colours appeared to the eye only at the boundaries of white and black, he recognized that a colour theory rooted in the properties of light must be inadequate. Goethe observed that there must be both light and dark, as well as a proximally positioned eye, in order for colour to arise. In one of his experiments, Goethe observed a card, half which was painted black, the other half white, using two orientations of the card. As Bortoft described, 'holding the prism so that it is oriented like the roof of a house turned upside down, with edges parallel to the boundaries [Goethe] look[ed] through the slanted side facing [him] toward the card'.[21] In the orientation with the black half on top, the colour spectrum appears in the white zone near the boundary, beginning with red, proceeding to orange, and ending with yellow farther from the boundary. In the orientation with the black half on the bottom, careful examination reveals that the colour spectrum appears in the black region at the boundary: first light blue, then dark blue, and finally purple the farthest away. Placing the prism next to the eye and observing a white wall yields nothing in the way of colour. Similarly, projecting a beam of white light through the prism and casting it on a wall some distance away, although produces a colour spectrum, produces a very different one described by Newton,[9] and an eye placed close to the prism is necessary to observe the phenomenon exactly as Goethe described it. Physically, this is due to the relatively shorter focal length that results from the proximal placement of the prism, as compared to Newton's experiment, in which the light refracted from a prism is cast onto a wall, some eighteen feet away. More importantly, this illustrates Goethe's dedication to the 'combination of subjective and objective experiments' to arrive at the whole of the phenomenon.[22] This was motivated by the goal of understanding the whole phenomenon of colour, including aspects examined by Newton, as well as those that are classified as subjective colour phenomena, such as colour mixing and after-images.[23] In this way, both light and eye were directly involved in the perception of colour.

---

[19] R. Steiner, *Goethe the Scientist*, New York, 1950, 1.

[20] H. Bortoft, *The Wholeness of Nature: Goethe's Way Toward a Science of Conscious Participation in Nature*, New York, 1996, 34.

[21] Bortoft, op. cit. (19), 40.

**Totalization: The Distal Prism Experiment**

Moreover, Goethe criticized Newton's experiment used to prove the second proposition of refrangibility of colour rays for the limitations it imposed on experimental variability, and consequently, upon the natural philosopher. Whereas Newton observed the colour spectrum cast on a wall at a fixed distance away from the prism, Goethe observed the cast spectrum on a white card which was progressively moved away from the prism.[24] When the card was close to the prism, analogous to the proximal positioning of the eye, the cast image was mostly white and circular, e.g., 'Refraction without the Appearance of Colour'.[23] As the card was moved away, the projected image elongated, gradually assuming an elliptical shape, and the coloured images became larger, finally merging at the center to produce green. Moving the card farther led to the increase in the size of the image, until finally the spectrum described by Newton in the *Opticks* was produced.[25] However, moving the card still farther led to the abrogation of the image, and the production of an oblong form consisting of violet, green, and orange. The image cast by the refracted beam was not fixed, but rather developed with increasing distance from the prism.[26] Consequently, Goethe saw the particular distance chosen by Newton to prove the second proposition of the *Opticks* as capriciously imposed.

---

[22] Goethe, op. cit. (3), 147.

[23] Goethe, op. cit. (3), liii.

In contrast to Newton, Goethe advanced a natural philosophy that sought to integrate all of the manifestations of a natural phenomenon. Thus, formulation of general laws regarding the function of natural phenomena was accomplished by the totalization of the phenomenal world of experience and experiment as to include all existent variations and consistencies of natural forms. For Goethe, it is by way of this integration that the inherently finite domain of human experience could mix with the infinity of Nature.

---

[24] Goethe, op. cit. (3), 127-38.

[25] It is interesting to note that this depiction of Newton's spectrum is slightly in error due to the appearance of colour blue, and is likely a result of Goethe's consultation with Erxleben's *Anfrangsgründe der Naturlehre*, instead of Newton's original description. D. Sepper, *Goethe, Newton, and Color: The Background and Rationale for an Unrealized Scientific Controversy*, (Ph. D. diss. AAT T-27972), The University of Chicago, 1981.

[26] N. Ribe, 'Goethe's Critique of Newton: A Reconsideration', *Studies in History and Philosophy of Science* (1985), 16, 315-35.

**A Gap: Knowledge Seeking and Intuitive Perception**

Goethe's colour theory was thus a phenomenology of colour, rather than an explanatory mechanism per se. Nevertheless, Goethe considered mechanics to be an intrinsic part of scientific inquiry. Consistent with such a view, he distinguished four types of cognitions, with their associated types of metaphysical organizations of human experience. Among the four were the knowledge seekers and the intuitively perceptive. The former were obliged to have a 'quiet, objective gaze, restless curiosity, and clear understanding'.[27] In this fashion, they were true Spinozists, striving after the objective reality and 'ideas which are true' in their equivalence with Nature.[28] On the other hand, the latter were quite Romantic, defined by their subjective faculties, and accessing Nature through their sense-impressions. The duality that existed between the objective and subjective components of truth, as polarized by these metaphysical distinctions, was bridged by Goethe's genetic method, making the 'relationship of the two [cognition types] clear and useful'.[29] Thus, Goethe concluded:

> If I look at the created object, inquire into its creation, and follow this process back as far as I can, I will find a series of steps. Since these are not actually seen together before me, I must visualize them in my memory so that they form a certain ideal whole. At first I will tend to think in terms of steps, but nature leaves no gaps, and thus, in the end, I will have to see this progression of uninterrupted activity as a whole. I can do so by dissolving the particular without destroying the impression itself'.[28]

---

[27] Goethe, op. cit. (5), 74.

[28] B. de Spinoza, *Correspondence*, (tr. R. H. M. Elwes), New York, 1955, 361.

[29] Goethe, op. cit. (5), 75.

In this way, Goethe's science existed on the 'borderline' of objective 'knowledge seeking' and subjective 'intuitive perception'.[26] In fact, Goethe asserted that strict Spinozism was not fruitful, as the 'seekers of knowledge,' who look at Nature with a quiet, objective gaze 'may cross themselves and bless themselves against imagination as often as they wish—before they know it, they will have to call on imagination's creative power to help'.[26]

**Methodology: Genetic and Comparative Methods**

This interest in the phenomenon's genesis and mechanism was transient and valuable only heuristically. The phenomenon's creation was understood by using the genetic and comparative methods, in the process of synthesizing the various sense-impressions of the object, but ultimately the ontogeny was made implicit, as the whole of the phenomenon was generated ideally. This genetic submergence was another expression of Goethe's Spinozism, since the phenomenon's genesis cannot be seen in Nature, as 'nature leaves no gaps.'[28] The creation, therefore, although made explicit during the mediation between object and subject, must ultimately be made implicit in imitation of Nature. Upon the conclusion of the genetic method, one was left with the 'ideal whole,' the form of the object's archetype.

The content of the archetype was obtained by using the comparative method, which 'teaches us what parts are common'.[30] This was accomplished by observing the phenomenon in a wide variety of its natural objectifications, comparing these observations to arrive at the invariant features and to synthesize them together. Although the stable features of the archetype were obtained by using the comparative method, synthesizing the plurality of empirical observations—'the idea [genetic form] must govern the whole'—and the genetic method provided the organization of the archetypal features obtained with the comparative method.[29] The whole of Goethe's scientific approach was made up by the combination of the genetic and comparative methods, with the latter providing the objective component, empirically complementing the subjective part which idealized to form the archetypal whole.

---

[30] Goethe, op. cit. (5), 118.

Goethe's way of observing required one to look into an object, and see behind one's intuition: 'to arrange things in order is a large and difficult undertaking [and] requires far more than sensory observation and memory'.[31] The missing link was 'insight into its [object's] character [and] striving of the human spirit'.[30] This mode of observation stretched from the phenomenal world into the noumenal one, in contrast to the Kantian assertion of the mediation of noumena by sense-impressions. Such reaching towards the objective was not strictly Spinozist either, since Goethe's encounter with the object became internalized and imagined, in stark contrast to Spinoza's view of imagination as a corruptor of true knowledge. The visualization of Goethe's genetic method thus involved seeing behind one's intuition and extending into the noumenal world, reversing the process, and imagining as the objective world stretched back across the subject-object boundary.

By means of this 'exact sensorial imagination,' the natural philosopher visualized the observation of the phenomenon, e.g., experimenting with the prism, and integrated all the various aspects of the phenomenon. Thus, the natural philosopher 'recreated in the wake of ever-creating nature'.[32] The goal of this process was to form an interjective organ of perception which could provide the natural philosopher with a deeper intuition of an object—yielding intuition of its essence and its organization—a goal impossible to achieve by simply observing an object or merely contemplating over it. In this way, the higher phenomenon did not appear to the senses. Instead, it was discovered within the sensory.

---

[31] Goethe, op. cit. (5), 73.

In his 1853 lecture 'On Goethe's Scientific Researches,' Hermann von Helmholtz asserted hat Goethe's 'theory of colours [was] an attempt to save the immediate truth of sense impression from the attacks of science'.[33] While it was true that Goethe's epistemology assigned critical importance to the role of intuition, Goethe was far more concerned with the mediate value of sense impressions. In the *Theory of Colours*, he directly rejected the idea that it was possible to have perceptions without theoretical constructions:

> Every act of seeing leads to consideration, consideration to reflection, reflection to combination, and thus it may be said that in every attentive look on nature we already theorise. But in order to guard against the possible abuse of this abstract view, in order that the practical deductions we look to should be really useful, we should theorise without forgetting that we are so doing, we should theorise with mental self-possession, and, to use a bold word, with irony.[34]

In this way, we were to impose our concepts onto Nature, but as natural philosophers we were to pre-empt this capacity. Thus, Goethe's scientific method did not divorce understanding of the phenomenon from the empirical reality of the phenomenon itself. By using the exact sensorial imagination, formulation of a hypothesis allowed the natural philosopher to cross from the world of empirical experience into the world of imagined concepts and back, in a delicate balance of precept and concept, but not localized to either one exclusively.

---

[32] Bortoft, op. cit. (19), 42.

[33] Helmholtz, op. cit. (2), 12.

[34] Goethe, op. cit. (3), xl-xli.

Thus, Goethe conceived three types of colour, characterized by varying degrees of the role of the subject in their experience, as reflected in the temporality of their perception.[35] By bridging the distinctions among these different colour types, and intuiting them in an 'unbroken series,' the natural philosopher thus developed a non-reductionistic view of the whole phenomenon.[36] The form of the recreated unity as obtained by using the genetic method did not reside in Nature, but persisted only in the natural philosopher's imagination during the synthetic process. As such, this was a fundamental basis for Goethe's discontent with Newtonian mechanics. Imaginative recreation of natural processes included a rational and mechanistic re-tracing of the process. But it recapitulated the unconscious purposiveness of Nature, in which the natural philosopher saw its telos, without bestowing on it a telic basis.

**On Goethe's Interjective metaphysics and Its Hybrid Nature**

The relationship between Kant's third critique,[37] and specifically its prescriptions for the use of teleological judgment in biology, and Goethe's phenomenalism and its use of teleology, are notable. Indeed, Goethe followed the Kantian course, conceiving final causes in Nature, but attributing no epistemic relevance to them as such, and thereby ascribing no telic functions to Nature. Goethe's metaphysics, however, although Kantian in some ways, was quite distinct in others. Specifically, Goethe uniquely hybridized Spinoza's and Kant's views, in such a way as to avoid the crisis of duality inherent in Kantian metaphysics, and the inadequacy of Spinoza's exclusive attention to the object and the dismissal of the productive role of imagination in the generation of knowledge. As such, Goethe was neither a strict Kantian, nor a true Spinozist.

---

[35] Goethe conceived of three types of colour: physiological, physical, and chemical; as originating in the eye, colorless media, and particular substances, respectively. Such a classification was based on the permanence of the perception of colour: fleeting, passing, and permanent, respectively. Goethe, op. cit. (3), lv.

[36] Goethe, op. cit. (3), lvi.

[37] I. Kant, *Critique of Judgment*, (trans. J. C. Meredith), Oxford, 1952.

Goethe began his Preface to the 1810 edition of the *Theory of Colours* by asserting that 'it [was] useless to attempt to express the nature of a thing abstractedly'.[38] He then added: 'Effects we can perceive, and a complete history of these effects would sufficiently define the nature of the thing itself.' Goethe was exposed to the Kantian contradiction inherent in such a view by Schiller, but failed to understand its metaphysical significance.[39] Nevertheless, two somewhat opposing aspects of this assertion are notable: i) that phenomena were discussed in terms of their effects on us; and ii) that such a phenomenal description could yield the nature of things in themselves, independent of our intuition of them. The Kantian distinction between object and subject, between noumena and phenomena, and between outer and inner worlds was thus problematized in terms of the gap that exists between these entities, and in terms of its impassability in Kantian metaphysics.

---

[38] Goethe, op. cit. (3), xxxvii.

[39] R. J. Richards, *The Romantic Conception of Life: Science and Philosophy in the Age of Goethe*, Chicago, 2002, 421-34.

Respecting his Spinozistic dispositions and their Naturism, Goethe set absolute priority to Nature and the objective world, and the influence which these had over the natural philosopher and his subjective faculties. 'Empirical evidence carries (and should carry) the greatest weight,' Goethe wrote, as Nature was 'a source of creation from the deepest depths to the loftiest heights'.[40] However, in this endeavor, Goethe recognized the role of the subject. Thus, Nature's essence was obtained by way of the imagination, which allowed the natural philosopher to imaginatively recreate natural processes, integrating their diverse phenetic qualities with the implicit understanding of their mechanism. However, this was done without losing sight of the primacy of Nature in determining the content of scientific theories. This use of the imagination was somewhat analogous to the Kantian imaginative syntheses, and the synthesis of apperception in particular,[41] but it was in stark contrast to the role of imagination in Spinoza's theory of knowledge,[42] which, if involved, in fact led to the formulation of false ideas.[27]

Although the role of the imagination and the subjective faculties of the mind in the ability to experience the world was quite Kantian, the fact that Goethean metaphysics allowed for the attainment of objective knowledge was not. This distinction was a source of disagreements between Schiller and Goethe, and was particularly manifest in Goethe's exclamation that he himself had 'ideas without knowing [them], and can even see them with [his] own eyes'.[36] The basic assumption of the Kantian view was that *a posteriori* elements in knowing the world were contingent on the particular *a priori* mode of our experience, and that only the pure elements were apodeictic. Although Goethe also considered that there were such *a priori* elements in our cognition, he characteristically distinguished his epistemology from Kant's by asserting that through experience, and consequently through intuition, it was possible to develop a way of knowing the world outside of the particulars of experience. As Goethe's archetypes were universally applicable, such an approach accessed the natural world not only phenomenally, but also noumenally.

---

[40] Goethe, op. cit. (5), 12.

[41] I. Kant, *Critique of Pure Reason*, (trans. N. K. Smith), New York, 1969.

[42] B. de Spinoza, *Ethics*, (trans. R. H. M. Elwes), New York, 1955.

Insofar as the outer world caused the inner world, this aspect of Goethe's thought was also consistent with Spinoza's views, as Goethe held that 'cause and effect must have a closer tie to one another than contiguity in space and time, [and that] they must be of a common qualitative kind'.[43] As Hegge noted, this view was similarly formulated by Spinoza: 'Things which have nothing in common cannot be understood, the one by means of the other'.[38] Indeed, such an integrated world-view was inherent in the integration of Nature and self that characterized Romantic thought.[44]

Thus, Goethe's metaphysics and epistemology were a hybrid of his Kantian and Spinozistic dispositions, positing the I interjectively at the interface of object and subject with a possibility of accessing things-in-themselves. In this light, Goethe's remark to Eckermann is particularly interesting: 'Kant never took any notice of me, although independently I was following a course similar to his'.[45] It can thus be suggested that

---

[43] H. Hegge, 'Theory of Science in the Light of Goethe's Science of Nature', *Inquiry* (1990), **15**, 363-86.

[44] Richards, op. cit. (36), 511-14.

[45] E. Cassirer, *Rousseau, Kant, and Goethe*, (tr. J. Gutmann, P. O. Kristeller, J. H. Randall, Jr.), 5 vols., Princeton, 1963, iii, 372.

Goethe's natural philosophy with its interjective moment succeeded at avoiding the crisis of post-Kantian metaphysics.[46]

**Goethe's Cosmology and the Pervasiveness of the Polar Archetype**

In addition to the totalization of science with respect to Nature, Goethe's *Theory of Colours* served another very important purpose. This goal reached to the opposite side of the object-subject interface, away from Nature, and towards Goethe, the natural philosopher. Specifically, Goethe saw the phenomenon of colour as a subject for the exegesis of his polar cosmology, employing colour on the basis of 'its close analogy [to the cosmological archetype] ... as equivalent [to it]'.[47]

Goethe's colour theory was another expression of the ultimate and highest archetype, that of unity of opposing forces. Introducing his *Theory of Colours*, Goethe wrote: 'to apply these designations [e.g., colour expressions of the cosmological archetype], this language of Nature to the subject we have undertaken; to enrich and amplify this language by means of the theory of colours and the variety of their phenomena, and thus facilitate the communication of higher theoretical views [of this cosmology], was the principal aim of the present treatise'.[48] These higher theoretical views permeated and over-reached the totality of Nature, as 'with light poise and counterpoise, [She] oscillates within her prescribed limits: now as a simple repulsion and attraction, now as an upsparkling and vanishing light, as undulation in the air, as commotion in matter, as oxydation and deoxydation'.[49]

---

[46] J. Habermas, *Postmetaphysical Thinking*, (tr. W. Hohengarten), Cambridge, 1996, 10-57.

[47] Goethe, op. cit. (3), xxxix.

[48] Goethe, op. cit. (3), xxxix-xl.

In this way, Goethe described magnetism, colour, wave theory, mechanics, and chemistry as all being expressions of one internally balanced and counterbalanced action of Nature.[50] Goethe's *Maxims and Reflections* expressed the universality and the supremacy of the polar archetype even more explicitly:

> Genesis and decay, creation and destruction, birth and death, joy and pain, all are interwoven with equal effect and weight; thus even the most isolated event always presents itself as an image and metaphor for the most universal.[51]

The pro- and contra- polarity inherent in Goethe's episteme was particularly apparent in his morphological writings. In describing how the archetype's expression was subject to the constraints of nature, Goethe brought up an example of the evolution of water- and air-borne organisms, and compared the two in terms of the inverse play-off of features which are ultimately balanced to yield a harmonious species:

> Water has a marked bloating effect on bodies it surrounds, touches, or penetrates to some degree. Thus the body of the fish, and especially its flesh, is swollen in conformity with the laws of the element. According to the laws of the archetype, this swelling of the body must be followed by a contraction of the extremities or auxiliary organs, not to mention further limitations of other organs.

---

[49] Goethe, op. cit. (3), xxxviii-xxxix.

[50] Goethe's cosmology may have an intriguing relationship with Schelling's natural philosophy, which stood in opposition to the Newtonian view of matter constituted by static particles, and instead asserted that matter is an equilibrium of dynamic forces that engage in 'polar opposition' to one another. F. W. J. von Schelling, *Ideas for a Philosophy of Nature*, (tr. E. Harris, P. Heath), Cambridge, 1988, 9-42.

[51] Goethe, op. cit. (5), 304.

> The air, by absorbing water, has a drying effect. Hence the archetype developed in the air will be as inwardly dry as the air is pure and lacking in moisture, giving rise to a more or less lean bird. Enough material will be left over for the formative force to clothe flesh and bone in rich array, and outfit the auxiliary organs fully. What in the fish was used for flesh is here left over for the feathers.[52]

Here the expression of the polar archetype as the formative force was applied to the morphological development of fish and birds, so that the effect of the aqueous environment on the fish was opposed by the contraction of its extremities, and the drying effect of air on birds was inversely neutralized by production of feathers. The formative force thus pushed forward only to be met with the counterbalance of natural constraints to create a harmonious organism. The polar archetype was related to the pure phenomenon of the opposition between light and dark as well. Writing in *On Morphology*, Goethe speculated:

> Plants and animals in their least perfect [initial ontogenic] state are scarcely to be differentiated. Hardly perceptible to our senses, they are a pinpoint of life, mutable, or semimutable. Are these beginnings—determinable in either direction—destined to be transformed by light into plant, or darkness into animal?[53]

In addition to the polarity that existed between plants and animals, there also existed a polarity within:

> The lower position is occupied by the root which works into the earth, belongs to the moisture and to the darkness. The stem, the trunk, or whatever may serve in its place, strives upward in exactly the opposite direction, toward the sky, the light, and the air.[48]

---

[52] Goethe, op. cit. (5), 122.

[53] Goethe, op. cit. (5), 65.

Goethe recognized the mystical nature of this question, since he 'would not trust [himself] to answer no matter how well [he is] supported with relevant observations and analogies'.[48] However, the mere inception of this question demonstrated the depth and breadth of Goethe's polar cosmology, which over-reached the totality of natural phenomena, from plants to animals, in the form of light and dark.[54]

The cosmological polarity of plants and animals was further related to the polarity of freedom and necessity: 'Plants attain their final glory in the tree, enduring and rigid, while the animal does so in man by achieving the highest degree of mobility and freedom'.[48] This was the interplay of freedom and necessity, as playing a role in the development of the self, in the process of which growth was attained by the counterbalance between life's determinism, insofar as the self cannot escape the circumstances of its inception, and one's will, which allowed the self to cultivate itself through interaction with the world.[55] In this way, the life of a plant was polarized toward the determinism of growth, and that of a man toward the freedom present in it. All of Goethe's thought, be it with regard to the personal formation of the self, morphological development, or ultimately colour, was united by a single over-reaching archetype, the clearest expression of which was the phenomenon of colour, and the inherent opposition between and unity of light and dark in its production.

---

[54] For a discussion of the relationship between Goethe's science and political economy, see M. Jackson, 'Natural and Artificial Budgets: Accounting for Goethe's Economy of Nature', *Science in Context* (1994), **7**, 409-31.

[55] J. W. von Goethe, *From My Life: Poetry and Truth*, (tr. R. R. Heitner), 2 vols., Princeton, 1994, i, 17, 44-45, 173-174, 241.

**Conclusion: Reaching to Span the Poles of a Rainbow**

The polar archetype of opposition and attraction over-reached not only Goethe's natural philosophy, but also his philosophical influences, bridging Goethe's Spinozistic and Kantian dispositions, as well as their internal frameworks. Thus, Goethe's genetic method, which in addition to standing on the borderline of object and subject, uniting these opposing Kantian entities, did so by simultaneously negating and embracing Spinoza's views of imagination and objectivism, respectively.

Insofar as Mephistopheles represented Goethe's belief that the original Light had the most important function in the cosmos—to be the mediator between 'spirit' and 'matter—' Goethe's theory of colour assumed a monumental importance.[56] As Pendlebury wrote: 'Newton set up a theory of light and incidentally of colour, which was apparently so conclusive, so successful, that it tended to exclude further investigation of the phenomena out of which the theory was carefully (and selectively) built up'.[57] This suppression of the creative and learning process—the cessation of growth—was precisely what incited Goethe to take such an opposing stance against Newton's wave theory.

Goethe saw in the popularity and dominance of Newtonian reductionism not only its explicit imposition on and irreverence of Nature, but more importantly, an actual act of imprisonment of the will, the same will to which he referred as part of the 'missing link' in relating to Nature. Thus, with the *Theory of Colours* Goethe set out to free this noble will, as one would free a 'princess' from a tower,[54] and render man free, not ontologically as Schiller, Fichte, and the other Jena Romantics surrounding Goethe sought to do in a post-Kantian world, but developmentally, allowing the natural philosopher to mediate his necessity and circumstance to explore the infinity within himself as well as that in Nature.

---

[56] J. W. von Goethe, *Faust*, (tr. D. Luke), Oxford, 1987, 42.

[57] D. Pendlebury, 'The Scientific Activity of J. W. von Goethe', *Systematics* (1965), **3**, 93-123.

Since according to Spinoza, infinity in Nature could not be perceived whole, Goethe's natural philosophy sought to unite its diverse forms and transcend the gap that existed between percept and concept. As Heller noted, Goethe himself reported in his diary an occasion on which he discussed this view of science with Schiller, particularly with respect to his morphological ideas, and insisted that 'perhaps there was still the possibility of another method, one that would not tackle Nature by merely dissecting and particularizing, but show her at work and alive, manifesting herself in her wholeness in every single part of her being'.[58]

Indeed, Goethe insisted that such a method could emerge from experience itself. As Goethe noted in the *Italian Journey*, one could perceive all of the particular plant forms as morphologically variable expressions of a universal plant principle that functioned in a telic manner in response to the constraints of different environments. Here the mixing and respective modification of Goethe's views of Spinoza's and Kant's epistemologies are most apparent. It is from the mixing of the intuitive and conceptual foundations of experience, that truthful, of the same kind of truth as in *Dichtung und Wahrheit*, science could be painted. While Spinoza pretended to do without hypotheses, dismissing the productive role of the imagination altogether, Kant favored the subsumption of the particular and uniquely situated experience of the natural world under the universal cognitive conditions that, albeit differently from the Spinozistic distortion, just as critically warp the work of a Goethean natural philosopher who seeks to understand the general principles of organization of the natural world. In contrast, Goethe, who in Werther's complementarily polarized yellow coat and blue frock,[59] experienced Nature interjectively, in the duality of object and subject, at the interface of percept and concept—bordering light and dark—and creating colour.

---

[58] E. Heller, *The Disinherited Mind*, New York, 1975, 6.

---

[59] J. W. von Goethe, *The Sorrows of Young Werther*, (tr. B. Pike), New York, 2004.